\documentclass[pre,aps,twocolumn, superscriptaddress]{revtex4}
\usepackage{amsmath}
\usepackage{epsfig}
\usepackage{amssymb}
\usepackage{color}

\begin{document}

\title{On the connection between dissipative particle dynamics and the
It\^{o}-Stratonovich dilemma}

\author{Oded Farago} 
\affiliation{Department of Biomedical
Engineering, Ben-Gurion University of the Negev, Be'er Sheva 85105,
Israel} 
\affiliation{Ilae Katz Institute for Nanoscale Science and
Technology, Ben-Gurion University of the Negev, Be'er Sheva 85105,
Israel}

\author{Niels Gr\o nbech-Jensen} 
\affiliation{Department of Mechanical
and Aerospace Engineering, University of California, Davis, California
95616, USA} 
\affiliation{Department Mathematics, University of
California, Davis, California 95616, USA}

\begin{abstract}

Dissipative Particle Dynamics (DPD) is a popular simulation model for
investigating hydrodynamic behavior of systems with non-negligible
equilibrium thermal fluctuations. DPD employs soft core repulsive
interactions between the system particles, thus allowing them to
overlap. This supposedly permits relatively large integration time
steps, which is an important feature for simulations on large temporal
scales. In practice, however, an increase in the integration time step
leads to increasingly larger systematic errors in the sampling
statistics. Here, we demonstrate that the prime origin of these
systematic errors is the multiplicative nature of the thermal noise
term in Langevin's equation; i.e., the fact that it depends on the
instantaneous coordinates of the particles. This lead to an ambiguity
in the interpretation of the stochastic differential Langevin
equation, known as the It\^{o}-Stratonovich dilemma. Based on insights
from previous studies of the dilemma, we propose a novel algorithm for
DPD simulations exhibiting almost an order of magnitude improvement in
accuracy, and nearly twice the efficiency of commonly used DPD
Langevin thermostats.

\end{abstract} 

\maketitle

Dissipative Particle Dynamics (DPD) \cite{warren-rev} is a popular
method employed for simulations of diverse molecular systems
including, colloidal suspensions, liquid crystals, polymers, and
bilayer membranes \cite{dpd-app1,dpd-app2,dpd-app3,dpd-app4}. It is
particularly appealing for investigations of multiscale phenomena
since, typically, the DPD particles do not represent individual atoms,
but rather coarse-grained (CG) collections of molecules. DPD was
originally introduced by Hoogerbrugge and Koelman \cite{originaldpd}
as a method for simulating hydrodynamic phenomena in complex
fluids. Specifically, the method targets {\em fluctuating
hydrodynamics}\/, i.e., hydrodynamics at the mesoscopic scales where
thermal fluctuations are important \cite{fluchyd,fluchyd2}. This is
achieved in DPD simulations by considering Langevin Dynamics (LD),
where each particle experiences a conservative forces and,
additionally, friction and random thermal forces that satisfy the
fluctuation-dissipation theorem \cite{espanol}. However, in {\em
conventional}\/ LD \cite{langevinbook}, the dissipative force acting
on the particle is given by $\vec{f}^D=-\gamma\vec{v}$, where
$\vec{v}$ is the velocity of the particle and $\gamma>0$ is a constant
friction coefficient. The random force is given by
$\vec{f}^R=\sqrt{2k_BT\gamma}\vec{R}(t)$, where $k_B$ is Boltzmann's
constant, $T$ is the temperature, and $\vec{R}(t)$ is a
Gaussian-distributed white noise with vanishing mean $\langle
R(t)_{\alpha}\rangle=0$ and memory-less auto-correlation $\langle
R(t)_{\alpha}R(t^\prime)_{\beta}\rangle=\delta(t-t^\prime)\delta_{\alpha\beta}$
($\alpha$ and $\beta$ denote Cartesian coordinates). In contrast, in
DPD the friction and random forces act in a pairwise fashion, and are
directed along the line connecting the centers of the particles. This
ensures that the total momentum of the system is conserved, which is
essential in order to capture the correct hydrodynamic behavior of
fluids at large scales \cite{Firsov}. Explicitly, the friction force
associated with each pair of particles $i\neq j$ is given by
\begin{eqnarray}
\vec{f}^D_{ij}=-\gamma\omega\left(r_{ij}\right)\left(\hat{r}_{ij}\cdot
\vec{v}_{ij}\right)\hat{r}_{ij},
\label{eq:friction}
\end{eqnarray}
where $r_{ij}=|\vec{r}_i-\vec{r}_j|$ is the pair-distance,
$\hat{r}_{ij}=\left(\vec{r}_i-\vec{r}_j\right)/r_{ij}$, and
$\vec{v}_{ij}=\vec{v}_i-\vec{v}_j$ is the relative velocity of the
particles. The random force is given by
\begin{eqnarray}
\vec{f}^R_{ij}=\sqrt{2k_BT\gamma\omega\left(r_{ij}\right)}\theta_{ij}
\hat{r}_{ij}.
\label{eq:random}
\end{eqnarray}
where $\theta_{ij}$ is Gaussian white noise satisfying
$\langle\theta_{ij}(t)\rangle=0$ and
$\langle\theta_{ij}(t)\theta_{kl}(t^\prime)\rangle=\left(\delta_{ik}\delta_{jl}
+\delta_{il}\delta_{jk}\right)\delta(t-t^\prime)$. The friction and
stochastic forces are modulated with a weight function,
$\omega\left(r_{ij}\right)$, that determines their range, $r_c$. The
most commonly used weight function is \cite{groot}
\begin{eqnarray}
\omega_{ij}=\left\{\begin{array}{ll}
\left(1-\frac{r_{ij}}{r_c}\right)^2 & r_{ij}<r_c \\
0 & r_{ij}\geq r_c,
\end{array} \right.
\label{eq:weight}
\end{eqnarray}
The total friction and stochastic forces exerted on the $i$th particle
are given by $\vec{f}^D_i=\sum_{j\neq i}\vec{f}^D_{ij}$ and
$\vec{f}^R_i=\sum_{j\neq i}\vec{f}^R_{ij}$, respectively.

As noted above, the particles in DPD simulations often represent CG
groups of atoms and molecules. Coarse-graining is believed to lead to
effective soft repulsive pair potentials; but, obviously, one has to
keep in mind that due to the softening of the effective interaction
potential, some features of the simulated system may change
\cite{comment-soft}. The conservative force used in DPD simulations is
usually given by \cite{groot}
\begin{eqnarray}
\vec{f}^C_{ij}=\left\{\begin{array}{ll}
a_{ij}\left(1-\frac{r_{ij}}{r_c}\right)\hat{r}_{ij} & r_{ij}<r_c \\ 0
& r_{ij}\geq r_c,
\end{array} \right.
\label{eq:conservative}
\end{eqnarray}
where $a_{ij}=a_{ji}$ are parameters determining the strength of the
repulsion, and the range $r_c$ is the same as in
Eq.~(\ref{eq:weight}). The total conservative force acting on the
$i$th particle is $\vec{f}^C_i=\sum_{j\neq i}\vec{f}^C_{ij}$. One of
the frequently proclaimed advantages of DPD simulations is that the
soft-core pair potential allows the particles to overlap and,
therefore, permits relatively large integration time steps $dt$, which
speeds up the simulations.  In practice, however, it is known that all
integration methods for DPD exhibit increasing artificial changes in
the sampling statistics as the discretization time step is enlarged
\cite{nikunen}. Thus, despite the enhanced numerical stability limit
obtained by softening the potentials, this feature imposes severe
restrictions on the size of the allowed time steps and, moreover,
requires one to validate results in order to asses the statistical
errors.

In this paper, we present a novel DPD integrator that, in
comparison to other DPD integrators, shows considerably smaller errors
in the computed averages of configurational thermodynamic
quantities. The new integration method is based on an integrator
recently presented by the authors (G-JF Integrator) that exhibits
minimal systematic errors in the sampling statistics of conventional
LD simulations \cite{gjf1,gjf2}, and on insights gained by
implementing the integrator to study LD in systems with spatially
varying friction coefficients \cite{gjf3,gjf4}. In the latter case, an
ambiguity, known in the literature as the {\em It\^{o}-Stratonovich
dilemma}\/, arises about the integration of the stochastic noise term
in Langevin's equation \cite{langevinbook,dilemma}. DPD belongs to the
same class of problems of stochastic dynamics with multiplicative
(state-dependent) noise, and it poses several unique complications
that we address in what follows.

In order to understand the origin of the problems in numerical
integration of DPD, we start by writing Langevin's equation of motion
\begin{eqnarray}
m_id\vec{v}_i=\left(\vec{f}^C_i+\vec{f}^D_i+\vec{f}^R_i\right)dt,
\label{eq:langevin}
\end{eqnarray}
where $m_i$ is the mass of the $i$th particle. The G-JF integrator
preserves the fluctuation-dissipation theorem in discrete time by
using the {\em exact} relationships:
\begin{eqnarray} 
\!\!\!\!\!\!\int_{t_n}^{t_{n+1}}\!\!\!\vec{f}^D_i\!
dt\!\!\!&=&-\int_{t_n}^{t_{n+1}}\!\!\!\gamma \vec{v}_idt=-\gamma
\left(\vec{r}^{n+1}_i-\vec{r}^n_i\right)\label{eq:fluc-dis1} \\
\!\!\!\!\!\!\int_{t_n}^{t_{n+1}}\!\!\!\vec{f}^R_i\!
dt\!\!\!&=&\!\!\!\!\sqrt{2k_BT\gamma}\int_{t_n}^{t_{n+1}}\!\!\!\vec{R}_i(t)dt
\!=\!\!\sqrt{2k_BT\gamma
dt}\vec{R}_i^{n+1}\!,\label{eq:fluc-dis2}
\end{eqnarray}
where $t^{n+1}=t^n+dt$ denotes discrete time,
$\vec{r}_i^n=\vec{r}_i\left(t_n\right)$, and $\vec{R}_i^{n+1}$ is a
vector whose coordinates are Gaussian random numbers of zero mean and
unity variance. Combining Eqs.~(\ref{eq:fluc-dis1}) and
(\ref{eq:fluc-dis2}) with the (second order in $dt$) approximations
used in the derivation of the Verlet algorithm \cite{verlet} for
Molecular Dynamics simulations in microcanonical ensembles, one
arrives at the G-JF algorithm for constant-temperature LD simulations,
which (after some rearrangement) reads
\begin{eqnarray}
\vec{r}^{n+1}_i&=&\vec{r}^n_i+b\left(v^n_idt+\frac{dt^2}{2m_i}
\vec{f}^n_i+\frac{dt}{2m_i}\vec{R}^{n+1}_i\right)\label{eq:gjf1}\\
\vec{v}^{n+1}_i&=&a\vec{v}^n_i+\frac{dt}{2m_i}\left(a\vec{f}^n_i+
\vec{f}^{n+1}_i\right)+\frac{b}{m_i}\vec{R}^{n+1}_i,\label{eq:gjf2}
\end{eqnarray}
where $\vec{v}^n_i=\vec{v}_i\left(t_n\right)$,
$\vec{f}^n_i=\vec{f}^C_i\left(t_n\right)$ is the {\em conservative}\/
force, and the constants 
\begin{eqnarray} 
b=\left(1+\frac{\gamma dt}{2m}\right)^{-1}\ ; \ a=\left(1-\frac{\gamma dt}{2m}
\right)b.
\label{eq:ab}
\end{eqnarray}
It has been demonstrated that, unlike other integrators, the G-JF
algorithm exhibits minimal changes in the configurational sampling
statistics as $dt$ is varied, up to the stability limit of the
integrator \cite{gjf1,gjf2,gjf5}.

When $\gamma=\gamma(\vec{r}_i)$ depends on the coordinate of the
particle, one needs to specify where along the path from
$\vec{r}^{n}_i$ to $\vec{r}^{n+1}_i$, the friction coefficient used in
Eqs.~(\ref{eq:fluc-dis2}) and (\ref{eq:ab}) is evaluated. This
ambiguity leads to the problem known as the It\^{o}-Stratonovich
dilemma, after the conventions $\gamma^{n+1}_{i{\rm
(I)}}=\gamma\left(\vec{r}^{n}_i\right)$ of It\^{o} \cite{ito}, and
$\gamma^{n+1}_{i{\rm (S)}}=\left[\gamma\left(\vec{r}^{n}_i\right)
+\gamma\left(\vec{r}^{n+1}_i\right)\right]/2$ of Stratonovich
\cite{stratonovich}. The vast majority of the literature on this topic
focuses on the overdamped (strictly non-inertial) limit of Langevin's
equation [$m_i\equiv0$ in Eq.~(\ref{eq:langevin})], where different
conventions lead to different statistical ensembles even for
infinitesimally small integration steps, $dt\rightarrow 0$. In the
case of full (inertial) dynamics (i.e., when the l.h.s.~of Langevin's
equation does {\em not}\/ vanish completely), the different
conventions lead to the same statistical sampling when $dt\rightarrow
0$. However, in numerical simulations with non-vanishing time steps, the
error caused by employing different conventions varies considerably
from one choice to another, and this error adds to the general error
caused by the integrator itself. In a previous study \cite{gjf3,gjf4},
we used the accurate G-JF integrator to study LD of a single
particle in a medium with space-dependent friction coefficient. We
demonstrated that both It\^{o} and Stratonovich interpretations lead
to noticeable deviations that scale linearly with $dt$ from the
equilibrium Boltzmann distribution. We proposed a new interpretation
that produces markedly smaller discrepancies between the computed and
the correct distributions and which, moreover, shows very little
sensitivity to $dt$ (and, thus, enables larger integration time
steps). The newly proposed convention for choosing the value of
$\gamma^{n+1}_i$ is based on the recognition that the random collision
forces between the Brownian particle and the moleculues of the heat
bath (which are not accounted for explicitly at molecular resolution)
are decomposed in Langevin's equation into two contributions. The
friction term represents the mean change in the momentum of the
particle due to the collisions, while the noise accounts for the
Gaussian statistical fluctuations around the mean value
\cite{gillespie}. We, therefore, consider the {\em deterministic}\/
part of Langevin's equation without the random component, and define
$\vec{s}^{n+1}_i=\vec{r}^n_i+\vec{v}^n_idt
+\left[\vec{f}^n_i-\gamma\left(\vec{r}^{n}_i\right)\vec{v}^n_i\right]
\left(dt^2/2m_i\right)$ that satisfies $\langle
\vec{r}^{n+1}_i\rangle\simeq\vec{s}^{n+1}_i+{\cal
O}\left(dt^3\right)$. Our new convention for $\gamma^{n+1}_i$ reads:
$\gamma^{n+1}_{i{\rm (G-JF)}}=\left[\gamma\left(\vec{r}^{n}_i\right)
+\gamma\left(\vec{s}^{n+1}_i\right)\right]/2$. This definition
resembles the Stratonovich interpretation for $\gamma^{n+1}_i$; yet it
does not create {\em spurious drift}. 
For a detailed discussion on the spurious drift problem, we
refer to refs.~\cite{gjf3,gjf4,lau,sancho} (and references
therein). In short, the fact that the noise term in
Eq.~(\ref{eq:langevin}) generates the distribution of momentum changes
around the mean value, implies that the r.h.s.~of
Eq.~(\ref{eq:fluc-dis2}) must satisfy
\begin{eqnarray}
\langle
\sqrt{2k_BT\gamma^{n+1}_idt}\,\vec{R}_i^{n+1}\rangle=0.
\label{eq:condition}
\end{eqnarray}
However, because $\vec{R}_i^{n+1}$ is a Gaussian random number with
zero mean, condition (\ref{eq:condition}) can only be fulfilled if
$\gamma^{n+1}_i$ and $\vec{R}_i^{n+1}$ are independent of each other,
which is {\em not}\/ the case with the seemingly physical Stratonovich
interpretation, where $\gamma^{n+1}_i$ depends on $\vec{r}^{n+1}_i$,
which itself depends on $\vec{R}_i^{n+1}$. It\^{o}'s interpretation
satisfies Eq.~(\ref{eq:condition}); however, it uses a poor estimation
for $\gamma^{n+1}_i$ (the initial value - completely ignoring the path
of the particle) and, therefore, also fails to produce accurate
statistical sampling for large $dt$. The new G-JF interpretation
satisfies condition (\ref{eq:condition}) (like It\^{o}), but with a
value of $\gamma^{n+1}_i$ representing a spatial average over the
ensemble of trajectories of the particle during the time step (like
Stratonovich).

DPD simulations present an even more challenging task of handling
multiplicative (state-dependent) noise. The complexity is mainly
linked to the fact that friction and noise forces act in a pairwise
fashion, and that they depend on both the relative coordinates and
velocities of the particles. Nevertheless, a considerably improved DPD
integrator can be devised, based on insights gained from our previous
investigations of the It\^{o}-Stratonovich dilemma. In order for the
fluctuation-dissipation theorem to be implemented appropriately in
discrete time, it is necessary to ensure that the friction and noise
forces associated with each pair act along the same direction, and
they must be weighted in a manner that on the one hand represents an
average over the time step ({\it a-la} Stratonivich convention), but
on the other hand independent of the random noise (in order to avoid
spurious drift, {\it a-la} It\^{o}). We, therefore, start by advancing
the system {\em without random forces}\/, which gives the
deterministic estimations for the new coordinates, $\vec{s}_i^{n+1}$,
and then compute the averages
\begin{eqnarray}
\!\!\!\!\!\!\!\!\vec{s}_i^{n+1/2}\!\!\!&\equiv&
\left(\vec{r}_i^n+\vec{s}_i^{n+1}\right)/2\nonumber \\
&=&\!
\vec{r}_i^n+\vec{v}_i^n\frac{dt}{2}+\left[\vec{f}^C_i\left(\vec{r}^n
\right)+\vec{f}^D_i\left(\vec{r}^n,\vec{v}^n\right)\right]\frac{dt^2}{4m_i},
\label{eq:shalf}
\end{eqnarray}
where $\vec{r}$ and $\vec{v}$ (to be distinguished from $\vec{r}_i$
and $\vec{v}_i$) denote dependence on coordinates and velocities of
all the particles. The coordinates $\vec{s}^{n+1/2}$ define the
directions of the friction and noise forces within the time step, as
well as the values of the friction coefficients. We, thus, continue
with calculating the random forces acting on the particles
\begin{eqnarray}
\!\!\!\vec{f}_i^{R^{n+1/2}}\!\!dt=\!\sum_{i\neq j}\!\!
\sqrt{2k_BT\gamma\omega\left(s_{ij}^{n+1/2}\right)dt}\,\,\theta_{ij}^{n+1}
\hat{s}_{ij}^{n+1/2}\!,
\label{eq:random2}
\end{eqnarray}
where $s_{ij}=|\vec{s}_i-\vec{s}_j|$ and
$\hat{s}_{ij}=\left(\vec{s}_i-\vec{s}_j\right)/s_{ij}$. We note the
following important technical point: In order to avoid the necessity
of recalculating the list of interacting particles associated with the
coordinates $\vec{s}_i^{n+1/2}$, we perform the summation in
Eq.~(\ref{eq:random2}) [as well as in Eqs.~(\ref{eq:friction2}) and
(\ref{eq:friction3}) below] over the list of interacting pairs
corresponding to $\vec{r}_i^n$. This excludes from the summation the
pairs with $s_{ij}^{n+1/2}<r_c$, for which $r_{ij}^n\geq r_c$. The
fraction of such pairs diminishes with $dt$ and their contribution to
the friction and noise forces is, anyhow, small. We have tested and
verified that including them in the sum has, indeed, almost no effect
on the computational results.

The calculation of the friction force poses a problem unique to DPD
simulations. In conventional LD, the impulse of the friction force on
each particle can be related to the displacement of the same particle
[see Eq.~(\ref{eq:fluc-dis1})], while in DPD the displacements of all
the particles are coupled. This precludes us from following the route
leading to Eqs.~(\ref{eq:gjf1}) and (\ref{eq:gjf2}), and enforces the
approximation of defining the velocity
\begin{eqnarray}
\!\!\!\!\vec{u}_i^{n+1/2}\!\!\equiv\!
\vec{v}_i^n\!+\!\!\left[\vec{f}^C_i\left(\vec{r}^n\right)
+\vec{f}^D_i\left(\vec{r}^n,\vec{v}^n\right)+\vec{f}_i^{R^{n+1/2}}\right]
\!\frac{dt}{2m_i},
\label{eq:uhalf}
\end{eqnarray}
and the associated friction forces
\begin{eqnarray}
\vec{f}_i^{D^{n+1/2}}\!\!=\!-\!\sum_{i\neq j}\!\!
\gamma\omega\!\left(s_{ij}^{n+1/2}\right)\!\left(\hat{s}_{ij}^{n+1/2}\!\!\cdot
\vec{u}_{ij}^{n+1/2}\right)\!\hat{s}_{ij}^{n+1/2},
\label{eq:friction2}
\end{eqnarray}
where $\vec{u}_{ij}=\vec{u}_i-\vec{u}_j$. The new
coordinates of the particles can now be computed using
\begin{eqnarray}
\!\!\!\vec{r}_i^{n+1}\!=\!
\vec{r}_i^n+\vec{v}_i^ndt\!+\!\!\left[\vec{f}^C_i\left(\vec{r}^n
\right)\!+\!\vec{f}_i^{D^{n+1/2}}\!\!+\!\vec{f}_i^{R^{n+1/2}}\right]
\!\frac{dt^2}{2m_i}
\label{eq:newt}.
\end{eqnarray}
Once the new coordinates are determined, we can calculate the change
in the relative coordinates
$\vec{\delta}_{ij}^{n+1/2}=\left(\vec{r}_i^{n+1}-\vec{r}_j^{n+1}\right)
-\left(\vec{r}_i^{n}-\vec{r}_j^n\right)$, and the associated
velocities $\vec{w}_{ij}^{n+1/2}\equiv\vec{\delta}_{ij}^{n+1/2}/dt$, and
replace approximation (\ref {eq:friction2}) with
\begin{eqnarray}
\vec{f}_i^{D^{n+1/2}}\!\!=\!-\!\sum_{i\neq j}\!\!
\gamma\omega\!\left(s_{ij}^{n+1/2}\right)\!\left(\hat{s}_{ij}^{n+1/2}\!\!\cdot
\vec{w}_{ij}^{n+1/2}\right)\!\hat{s}_{ij}^{n+1/2}.
\label{eq:friction3}
\end{eqnarray}
We also compute the new deterministic forces,
$\vec{f}_i^C\left(\vec{r}^{n+1}\right)$, and then evaluate the new
velocities via
\begin{eqnarray}
\!\!\!\vec{v}_i^{n+1}\!\!\!=\!
\vec{v}_i^n\!\!+\!\!\!\left[\frac{\vec{f}^C_i\left(\vec{r}^n
\right)\!+\!\!\vec{f}^C_i\left(\vec{r}^{n+1}
\right)}{2}\!+\!\vec{f}_i^{D^{n+1/2}}\!\!+\!\vec{f}_i^{R^{n+1/2}}\!\right]
\!\!dt.
\label{eq:newv}.
\end{eqnarray}
We ``close the loop'' by calculating the friction forces
$\vec{f}_i^D\left(\vec{r}^{n+1},\vec{v}^{n+1}\right)$ to be used at
the next application of Eq.~(\ref{eq:shalf}).

\begin{figure}[t]
   \centering\includegraphics[width=0.45\textwidth]{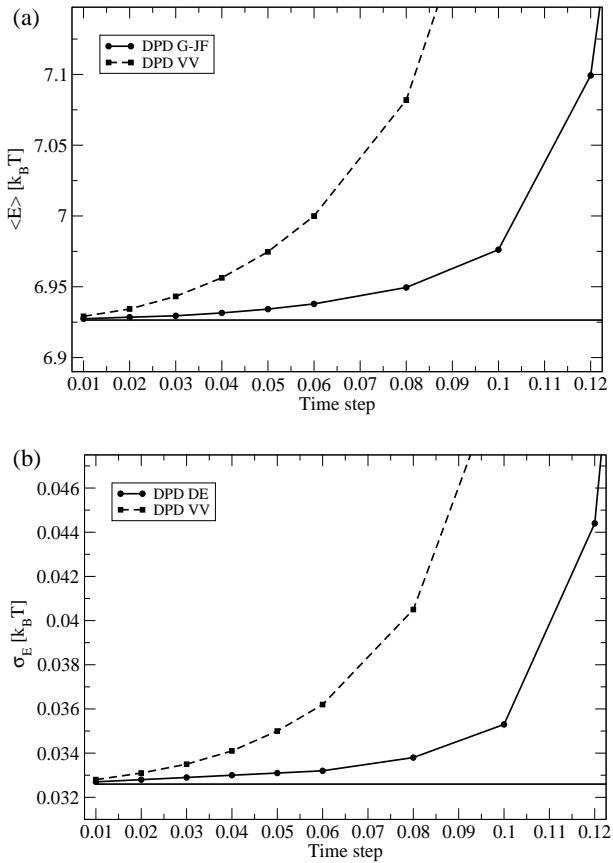} 
    \caption{Mean (a) and standard deviation (b) of the potential
      energy per particle, computed with different integration time
      steps. Results for the DPD-DE and DPD-VV methods are plotted in
      circles (with solid line) and squares (with dashed line),
      respectively. The lines serve as guides to the eye.}
    \label{fig:fig1}
\end{figure}

The sequence of Eqs.~(\ref{eq:shalf})-(\ref{eq:newv}) constitutes our
proposed new DPD integrator, which we term DPD-DE after the
``deterministic estimation'' of $\vec{s}_i^{n+1}$ in
Eq.~(\ref{eq:shalf}). To test the algorithm, we simulate a system of
$N=500$ identical particles in a cubic box of length $L=5$ with the
parameter set $r_c$=1, $k_BT=1$, $m_i=1$, $a_{ij}=25$, and
$\gamma=4.5$. This system, with the same set of parameters, has
recently been used in ref.~\cite{leimkuhler} for comparison between
several DPD integrators. As a benchmark, we use the DPD
Velocity-Verlet (DPD-VV) method of Besold {\em et al.}\/\cite{besold},
which is implemented in several popular simulation packages. We note
that in the simulations of the very same system in
ref.~\cite{leimkuhler}, the accuracy and efficiency of the DPD-VV
algorithm was found to be almost identical to other commonly used DPD
Langevin integrators such as Shardlow's splitting method
\cite{shardlow,avalos}. Therefore, the DPD-VV results also allow
comparison with other integration schemes for constant-temperature
DPD. We also note that methods for DPD simulations with energy
conservation exist (e.g., \cite{avalos2,ripoll}), but the discussion
of constant-energy DPD is beyond the scope of this paper. The
performance of the integrator is evaluated by measuring the mean and
standard deviation of the potential energy of the system in
simulations with increasing time steps. These quantities characterize
the quality of configurational sampling. For each time step, ranging
from $dt=0.01$ and up to a time step showing significant deviations
from the asymptotic $dt\rightarrow 0$ limit, we simulated the system
for $1.44\times 10^6$ time units, and sampled the energy at intervals
of $1.2$ time units. Our results for the mean ($\langle E\rangle$) and
standard deviation ($\sigma_E$) of the potential energy (normalized
per particle) are plotted, respectively, in Fig.~\ref{fig:fig1}. We
observe that both methods exhibit an increase in the measured $\langle
E\rangle$ and $\sigma_E$ with $dt$ indicating unwanted changes in the
sampling statistics.  However, per $dt$, the results of the DPD-DE
integrator of this work appear to be about 6-7 times more accurate
(i.e., exhibiting smaller relative errors) than the results of the
DPD-VV method. A similar degree of improvement in accuracy has been
found in simulations of both denser and more dilute systems, and for
different values of the the parameter $a_{ij}$ representing
stronger/weaker repulsion between the particles.

\begin{figure}[t]
   \centering\includegraphics[width=0.45\textwidth]{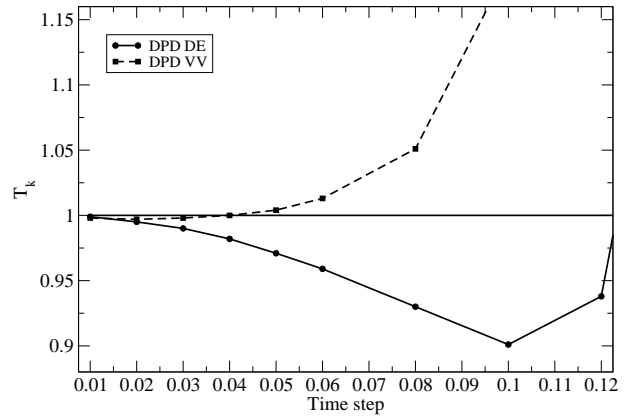} 
    \caption{The computed kinetic temperature as a function of the
    integration time steps. Results for the DPD-DE and DPD-VV methods
    are plotted in circles (with solid line) and squares (with dashed
    line), respectively. The lines serve as guides to the eye.}
    \label{fig:fig2}
\end{figure}

Fig.~\ref{fig:fig2} depicts the results for the simulated kinetic
temperature, $T_k=(2/3)\langle K\rangle/(N-1)$ (where $K$ is the
kinetic energy), as a function of $dt$. One may erroneously conclude
from the results for $T_k$ that DPD-VV performs better than
DPD-DE. This impression, however, is incorrect. It has been now well
established (see numerous discussions on this point in, e.g.,
\cite{gjf1,gjf2,gjf5,leimkuhler,eastwood}) that in contrast to the
potential energy, the simulated kinetic energy is not very important
and cannot be taken as a reliable measure for the accuracy of a
simulation method. This feature of numerical integrators does not
originate from the discretization of the friction and noise
forces. This is an inherent property of the classic Verlet algorithm
where the discrete-time momentum $\vec{v}_i^n$ is not exactly
conjugated to the coordinate $\vec{r}_i^n$. For this reason, one
should not attempt to use (with any integrator) quantities, such as
momentum autocorrelations, for precise measures, unless very small
integration time steps are applied.

\begin{figure}[t]
   \centering\includegraphics[width=0.45\textwidth]{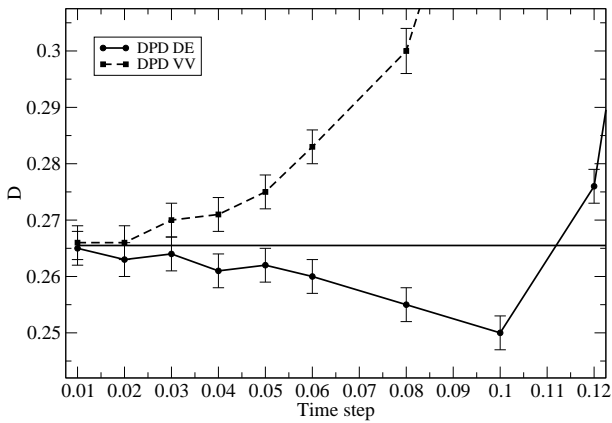} 
    \caption{The computed diffusion coefficient as a function of the
    integration time steps. Results for the DPD-DE and DPD-VV methods
    are plotted in circles (with solid line) and squares (with dashed
    line), respectively. The lines serve as guides to the eye.}
    \label{fig:fig3}
\end{figure}

To ensure that the new integrator is useful for DPD simulations, it is
necessary to also demonstrate that it produces the correct dynamics,
at least as accurately as other algorithms. As a measure for the
dynamical evolution of the system, we consider the diffusion
coefficient~\cite{nikunen:2003}
  \begin{eqnarray}
    D=\lim_{t\rightarrow\infty}\frac{1}{6Nt}\left\langle
    \sum_{i=1}^N\left[\vec{r}_i(t)-\vec{r}_i(0)\right]^2\right\rangle,
  \end{eqnarray}
whose value depends only on the discrete-time coordinates
$\left\{\vec{r}_i^n\right\}$, but not on the discrete-time momenta
$\left\{\vec{v}_i^n\right\}$. Results for $D$ as a function of $dt$
are shown in Fig.~\ref{fig:fig3}. As expected, the results of both
integrators converge to the same limit when $dt\rightarrow 0$,
indicating convergence to the correct dynamical
behavior. Interestingly, the trends in the variations of $D$ resembles
the trends in $T_k$ (Fig.~\ref{fig:fig2}). Also noticeable, the
discretization time errors of the DPD-DE algorithm are always smaller
the errors of the DPD-VV algorithm. The last observation suggests that
the new DPD-DE method improves not only configurational sampling, but
also provide a better dynamical description of DPD systems.

The improvement by a factor of 6-7 in configurational sampling
accuracy is outstanding considering that, per $dt$, all currently
available Langevin thermostats for DPD simulations exhibit relative
errors essentially identical to the one of the DPD-VV method
\cite{leimkuhler}. This property suggests that the main source of
numerical error in Langevin DPD thermostats is the application of
It\^{o}'s interpretation to the friction coefficients, which is the
common feature of all of these methods. The DPD-DE integrator of this
work uses a different convention, which is based on spatial averaging
of the friction along the trajectory that the particle would follow
had the random noise force been turned off \cite{comment1}. This new
convention differs from the {\em seemingly more physical}\/
Stratonovich convention that is based on the actual trajectory of the
particle, and which also takes into account the influence of the
random force along the trajectory. The Stratonovich interpretation
represents an incorrect reading of Langevin's differential
equation. In Langevin's equation, the friction force represents the
mean change in the momentum of a particle, while the noise term
accounts for the statistical distribution {\em around}\/ the mean
value. The Stratonivich interpretation ``mixes'' the two terms and,
therefore, it leads to spurious drift \cite{gjf3,gjf4}.

We close by noting that in order to asses the computational efficiency
of integrators, one also needs to take into account the CPU time
required to perform a single time step. For that purpose, we adopt the
criterion suggested in ref.~\cite{leimkuhler}, which defines the
numerical efficiency as the step size giving the same relative
accuracy as the DPD-VV method with step size $dt=0.05$, divided by the
CPU time. From Fig.~\ref{fig:fig1} we read that the DPD-DE method with
$dt=0.1$ has the same accuracy as the DPD-VV method with
$dt=0.05$. Simulations on several different machines also reveal that
the run time of DPD-DE is about 1.15-1.3 larger than that of DPD-VV
\cite{comment2}. Thus, the scaled efficiency of DPD-DE is about
$155\%-175\%$, placing it second in the list of integrators examined
in ref.~\cite{leimkuhler} in terms of computational efficiency, just
an inch behind the method that came first with scaled efficiency of
$187\%$. However, the latter method, as well as {\em all}\/ other
integration methods ranked at the top places of the list, are based on
a Nos\'{e}-Hoover thermostat. Such methods are more complicated for
implementation, and their optimization requires fine-tuning of
additional friction parameters. In contrast, DPD-DE is a {\em pure}\/
Langevin thermostat having only a single tunable friction parameter
$\gamma$ [see Eqs.~(\ref{eq:friction}) and (\ref{eq:random})].  It,
thus, offers both ease of implementation and benefit of accuracy.

This work was supported by the Israel Science Foundation (ISF), Grant
No. 1087/13, and by the U.S.~Department of Energy, Project
No. DE-NE0000536000.


\begin{thebibliography}{99}
\bibitem{comment1} For practical reasons, the algorithm is based on a
forward approximation of the trajectory.
\end{thebibliography}
\end{document}